\documentclass[12pt]{amsart}

\usepackage{amsmath}
\usepackage{amssymb}

\usepackage[utf8]{inputenc}
\usepackage[T1]{fontenc}

\newtheorem{theorem}{Theorem}[section]

\theoremstyle{definition}
\newtheorem{definition}{Definition}[section]

\numberwithin{equation}{section}

\begin{document}
	
	\title{Toeplitz Quantization of a 
	Free $ * $-Algebra}

	\author{Stephen Bruce Sontz}
	
	\address{Centro de Investigaci\'on en Matem\'aticas, A.C.  
		(CIMAT), 
		Jalisco S/N, Mineral de Valenciana, CP 36023, 
		Guanajuato, Mexico}
	\email{sontz@cimat.mx}
	
	\begin{abstract}
		In this note 
		we quantize 
		the free $ * $-algebra
		generated by finitely many variables,
		which is a new example of
	    the theory of Toeplitz quantization 
		of $ * $-algebras as developed 
		previously by the author.  
		This is achieved by defining Toeplitz 
		operators with symbols in that 
		non-commutative free $ * $-algebra. 
		These are densely defined operators 
		acting in a Hilbert space. 
		Then 
		creation and annihilation operators 
		are introduced as special cases of 
		Toeplitz operators, and their properties 
		are studied. 
		\begin{center}
			Dedication: 
			\\
        To Nikolai Vasilevski 
        in celebration of his 70th birthday. 
        \end{center}
			
	\end{abstract}
		
	\maketitle
	
	\noindent 
	{\bf Keywords:} 
	Toeplitz operators, 
	creation and annihilation operators 
	

\section{Introduction}

The basic reference for this paper is 
\cite{me} where a general theory 
of Toeplitz quantization of
$ * $-algebras is defined and studied. 
More details including motivation and 
references can 
be found in \cite{me}.

\section{The Free $ * $-algebra}

The  example in this paper 
is the free algebra on $2 n$ 
non-commuting variables 
$ \mathcal{A} = \mathbb{C} \{ \theta_1, \overline{\theta}_1, \dots , \theta_n, \overline{\theta}_n \}$. 
In particular, the variables 
$ \theta_j, \overline{\theta}_j $ do not commute 
for $ 1 \le j \le n $.  
The {\em holomorphic sub-algebra} is defined by  
$ \mathcal{P} := \mathbb{C} \{ \theta_1, \dots , \theta_n \}$, the free algebra on $n$ variables. 
The {\em $*$-operation} (or {\em conjugation}) on 
$\mathcal{A}$ is defined on the generators by
$$
\theta_j^* := \overline{\theta}_j \quad \mathrm{and} \quad \overline{\theta}_j{\,\! ^*} := \theta_j,
$$
where $j = 1, \dots, n$. 
This is then extended to finite products of these $2n$ elements in the unique way that 
will make $\mathcal{A}$ into a $*$-algebra with $1^* = 1$.
As explained in more detail in a moment these 
products form a vector space 
basis of $\mathcal{A}$, 
and so we extend the $*$-operation 
to finite linear combinations of them 
to make it an {\em anti-linear} 
map over the field $\mathbb{C}$ of complex numbers. 
Therefore, $\mathcal{P}$ is not a sub-$*$-algebra. 
Rather, we have $\mathcal{P} \cap \mathcal{P}^* = \mathbb{C} 1$. 
Moreover, $\mathcal{P}$ is a non-commutative sub-algebra of $\mathcal{A}$ if $ n \ge 2 $. 
This set-up 
easily generalizes to infinitely many pairs
of non-commuting variables 
$ \theta_j, \overline{\theta}_j $. 

The definition of $ \mathcal{P} $ is 
motivated as a non-commutative 
analogy to 
the commutative 
algebra of holomorphic polynomials in the 
Segal-Bargmann space 
$ L^{2} (\mathbb{C}^{n}, e^{-|z|^{2}} \mu_{Leb}) $, 
where $ \mu_{Leb} $ is  Lebesgue measure 
on the Euclidean space 
$ \mathbb{C}^{n} $. 
(See \cite{bargmann} and \cite{brian}.) 
This is one motivation behind using the notation 
$ \mathcal{P} $ for this sub-algebra. 

We will later introduce a projection operator 
$ P : \mathcal{A} \to \mathcal{P} $ using 
a sesqui-linear form defined on $ \mathcal{A} $. 
This is an essential ingredient in the following 
definition. 
\begin{definition}
Let $ g \in \mathcal{A} $ be given. 
Then we define the {\em Toeplitz operator} $ T_{g} $
with symbol $ g $ as 
$ T_{g} \, \phi := P (\phi g) $ for all 
$ \phi \in \mathcal{P} $. 
It follows that 
$ T_{g} : \mathcal{P} \to \mathcal{P} $ is
linear. 
We let 
$ \mathcal{L} (\mathcal{P}) :=
\{  T : \mathcal{P} \to \mathcal{P} \,|\, 
T \mathrm{~is~linear} \}$.  
Then the linear map 
$ \mathcal{A} \ni g \mapsto 
T_{g} \in \mathcal{L} (\mathcal{P}) $ is 
called the {\em Toeplitz quantization}.
\end{definition} 

Multiplying the symbol $ g $ on the left of 
$ \phi $ gives a similar theory, which we
will not expound on in further detail. 

The sesqui-linear form on $ \mathcal{A} $ when
restricted to $ \mathcal{P} $ will turn out to 
be an inner product. 
So $ \mathcal{P} $ will be a pre-Hilbert space
that is dense in its completion, denoted as 
$ \mathcal{H} $. 
This is another motivation for using the 
notation $ \mathcal{P} $ for this sub-algebra. 
So every Toeplitz operator $ T_{g} $ 
will be a densely defined linear operator acting
in the Hilbert space $ \mathcal{H} $. 

The definition of the sesqui-linear form 
on $\mathcal{A}$ is a more involved story. 
To start it we let $\mathcal{B}$ be the standard basis of $\mathcal{A}$ 
consisting of all finite \textit{words} 
(or \textit{monomials}) in the finite alphabet 
$ \{ \theta_1, \overline{\theta}_1, \dots , \theta_n, \overline{\theta}_n \}$, which 
has $2 n$ letters. 
The empty word (with zero letters) is 
the identity element $1 \in \mathcal{A}$. 
Let $f \in \mathcal{B}$ be a word in our alphabet. 
We let $l(f)$ denote the \textit{length} of $f$, that is, the number of letters in $f$. 
Therefore $l(f) = 0$ if and only if $f =1$. 
\begin{definition}
	Let $f \in \mathcal{B}$ with $l(f) > 0$. 
	Then we say that $f$ {\em begins with a $\theta$} if the first letter of $f$
	(as read from the left) is an element of $\{ \theta_1, \dots , \theta_n \}$; 
	otherwise, we say that $f$ {\em begins with a $\overline{\theta}$}. 
	
	If $f = 1$, then we say that $f$ 
    {\em begins with a $\theta$} and
	$f$ {\em begins with a $\overline{\theta}$}. 
\end{definition}
\noindent 
\textbf{Remark:} 
Suppose $l(f) > 0$ and that $f$ begins with a $\theta$. 
Then $f$ has a \textit{unique} representation as
\begin{equation}
\label{unique-rep-of-f}
f = \theta_{i_1} \cdots \theta_{i_r} \overline{\theta}_{j_1} \cdots  \overline{\theta}_{j_s} f^\prime,
\end{equation}
where $r \ge 1$, $s \ge 0$ and $f^\prime$ begins with a $\theta$. 
That is to say, the word $f$ begins with $r \ge 1$ occurrences of $\theta$'s followed by
$s \ge 0$ occurrences of $\overline{\theta}$'s and finally another word $f^\prime$ that begins
with a $\theta$. 
Note that if $s=0$, then $f^\prime = 1$. 
We also have that $l(f^\prime) < l(f)$. 
As a simple example of this representation, note that each basis element 
$f = 
\theta_{i_1} \cdots \theta_{i_r}$ in $\mathcal{P}$  
with $ r \ge 1 $
has this representation with 
$s=0$ and $f^\prime =1$. 

Dually, suppose that $l(f) > 0$ and that $f$ begins with a $\overline{\theta}$. 
Then $f$ has the obvious dual representation. 

Now we are going to define a sesqui-linear form 
$\langle f , g \rangle$ for $f, g \in \mathcal{A}$ 
by first defining it on 
pairs of elements of the basis $\mathcal{B}$ and extending sesqui-linearly, which for us
means anti-linear in the first entry and linear in the second. 
The definition on pairs will be by recursion on the length of the words. 
To start off the recursion for $l(f) = l(g) =0$ 
(that is, $ f = g =1 $)
we define 
$$
\langle f , g \rangle =
\langle 1 , 1 \rangle := 1.
$$
This choice is a convenient 
normalization convention. 

The next case we consider is $l(f) > 0$, $f$ begins with a $\theta$ and $g =1$. 
In that case using \eqref{unique-rep-of-f}
 we define recursively 
\begin{align*}
\langle f , 1 \rangle &= 
\langle  \theta_{i_1} \cdots \theta_{i_r} \overline{\theta}_{j_1} \cdots  \overline{\theta}_{j_s} f^\prime, 1 \rangle
\\
&:= w (i_1, \dots,  i_r) \delta_{r,s} \delta_{i_1, j_r} \cdots \delta_{i_r, j_1} \langle f^\prime , 1 \rangle 
\\
&= w(i)  \delta_{r,s} \delta_{i,j^T} \langle f^\prime , 1 \rangle
\end{align*}
where $r \ge 1$ and $w (i) \equiv w(i_1, \dots,  i_r) > 0$ is a positive weight. 
Here we also define the (variable length) multi-index $i=(i_1, \dots,  i_r)$ and
$j^T := (j_s, \dots , j_1)$ to be the reversed multi-index of the multi-index $j=(j_1, \dots , j_s)$. 
It follows that $\langle f , 1 \rangle \ne 0$ in this case implies that we necessarily have 
$\langle f^{\prime} , 1 \rangle \ne 0$ and 
$$
f = \theta_{i_1} \cdots \theta_{i_r} \overline{\theta}_{i_r} \cdots \overline{\theta}_{i_1} f^\prime.
$$

Moreover, by recursion $f^\prime$ must 
also has this same form as $f$. 
Since the lengths are strictly decreasing ($l(f) > l(f^\prime) > \cdots$), this recursion terminates
in a finite number of steps. 
Thus the previous equation can then be written using the obvious notations
$\theta_i := \theta_{i_1} \cdots \theta_{i_r}$ and 
$\overline{\theta}_{i^T} := \overline{\theta}_{i_r} \cdots \overline{\theta}_{i_1}$ as 
$$
f = \theta_i \overline{\theta}_{i^T} f^\prime. 
$$

Symmetrically, for $f = 1$, $l(g) > 0$ and $g$ begins with a $\theta$ 
we write 
$ g = \theta_{k_1} \cdots \theta_{k_t}
\overline{\theta}_{l_1} \cdots  
\overline{\theta}_{l_u} g^\prime $ 
uniquely so that $ t \ge 1 $ and 
$ g^{\prime} $ begins with a $ \theta $ 
and define recursively
\begin{align*}
\langle 1 , g \rangle &= 
\langle 1,   \theta_{k_1} \cdots \theta_{k_t} \overline{\theta}_{l_1} \cdots  \overline{\theta}_{l_u} g^\prime \rangle
\\
&:= w (k_1, \dots,  k_t) \delta_{t,u} \delta_{k_1, l_u} \cdots \delta_{k_t, l_1} \, \langle 1, g^\prime  \rangle 
\\
&= w(k) \delta_{t,u} \delta_{k, l^T} \, \langle 1, g^\prime  \rangle. 
\end{align*} 

Next suppose that $l(f) > 0$ and $l(g) > 0$ and that both $f$ and $g$ begin with a $\theta$ 
and are written as above.  
In that case, we define
\begin{align}
\label{both-lengths-positive}
\langle f , g \rangle &= \langle 
\theta_{i_1} \cdots \theta_{i_r} \overline{\theta}_{j_1} \cdots  \overline{\theta}_{j_s} f^\prime, \, 
\theta_{k_1} \cdots \theta_{k_t} \overline{\theta}_{l_1} \cdots  \overline{\theta}_{l_u} g^\prime \rangle 
\\
&:= w(i,l^T) \delta_{r+u, s+t} \delta_{ (i,l^T), (k,j^T) } \, \langle f^\prime , g^\prime \rangle,
\nonumber
\end{align} 
where $(i,l^T) := (i_1, \dots, i_r, l_u, \dots, l_1)$ is the concatenation of the two 
multi-indices $i$ 
and $l^T = (l_u, \dots, l_1)$. 
(Similarly for the notation $(k,j^T)$.) 

The definitions for two words that begin with a $\overline{\theta}$ are dual to these definitions. 
We use the same weight factors for this dual part, 
though new real weight factors could have been used. 

There is still one remaining case for which we have yet to define the sesqui-linear form.
That case is when $f$ begins with a $\theta$, $g$ begins with a $\overline{\theta}$, 
(or {\em vice versa}), 
$l(f) > 0$ 
and 
$l(g) > 0$. 
In that case we define $\langle f, g \rangle := 0$. 

\begin{theorem}
\label{thm-symmetry}
	The sesqui-linear form on $\mathcal{A}$ is 
	{\em complex symmetric}, that is, 
	$$
	\langle f, g \rangle ^* = \langle g ,f \rangle \quad \mathit{for~all~} f,g \in \mathcal{A}. 
	$$
\end{theorem}
\noindent
\textbf{Proof:}
The proof is by induction following the various cases 
of the recursive definition of the 
sesqui-linear form. 
First, for $l(f) = l(g) = 0$ we have $f=g=1$ in which case
$$
\langle 1, 1 \rangle ^* = 1 ^* = 1 = \langle 1, 1 \rangle.
$$
Next we take the case $l(f) > 0$, $f$ begins with a $\theta$ and $l(g) =0$.
Then we write $f = \theta_i \overline{\theta}_j f^\prime$ for multi-indices $ i,j $ 
of lengths $ r,s $ respectively 
and $ f^{\prime} $ begins with a $ \theta $. 
So we calculate 
\begin{equation*}
\langle f, 1 \rangle^* = \big( w(i) \delta_{r,s}
\delta_{i,j^T} \langle f^\prime , 1 \rangle \big)^*
= w(i) \delta_{r,s} \delta_{i,j^T} 
\langle 1, f^\prime \rangle, 
\end{equation*}
where we used the induction hypothesis and 
the reality of the weight $ w(i) $ for the last step.
On the other hand, we have by definition that
\begin{equation*}
\langle 1, f \rangle = \langle 1, \theta_i \overline{\theta}_j f^\prime \rangle 
= w(i) \delta_{r,s} \delta_{i,j^T} \langle 1, f^\prime \rangle. 
\end{equation*}
This proves that $\langle f, 1 \rangle^* = \langle 1, f \rangle$.
Similarly, one shows $\langle 1, g \rangle^* = \langle g , 1 \rangle$, where $ l(g) > 0 $ 
and $ g $ begins with a $ \theta $. 

For the case where $l(f) > 0$ and $l(g)> 0$ and both 
$ f  $ and $ g $ begin 
with a $\theta$, we write $f =  \theta_i \overline{\theta}_j f^\prime$ and
$g =  \theta_k \overline{\theta}_l g^\prime$, where
$ i,j,k,l $ are multi-indices of 
lengths $ r,s,t,u $ respectively 
and $ f^{\prime} , g^{\prime} $ begin 
with a $ \theta $. 
Then we see by induction that
\begin{align*}
\langle g , f \rangle^* &= 
\langle \theta_k \overline{\theta}_l g^\prime ,  \theta_i \overline{\theta}_j f^\prime \rangle^*
= \big( w(k, j^T) \, \delta_{t+s, u+r} \, \delta_{(k,j^T), (i,l^T) } \, \langle g^\prime , f^\prime \rangle \big)^*
\\
&= w(i, l^T) \, \delta_{r+u, s+t } \, \delta_{(i,l^T) , (k,j^T)} \, \langle f^\prime , g^\prime \rangle
= \langle f, g \rangle. 
\end{align*}
The proofs for words that begin with $\overline{\theta}$ are similar. 
The final case is if one of the pair 
$f,g$ begins with a $\theta$ and the other 
begins with a $\overline{\theta}$. 
But then $\langle f , g \rangle = 0$ as well as $\langle g , f \rangle =0$. 
So in this final case the identity 
is trivially true. 
$\quad \blacksquare$

\vskip 0.1cm 
\noindent
While the sesqui-linear form 
is complex symmetric according to this proposition, 
when $n \ge 2$ 
it does not satisfy the nice properties 
with respect to the $ * $-operation as were
given in \cite{me}. 
We recall that those properties are
\begin{align}
\label{comp-prop-1}
\langle f_1, f_2 g \rangle &= \langle f_1 g^* , f_2 \rangle, 
\\
\label{comp-prop-2}
\langle f_1, f_2 g \rangle &= \langle f_1  f_2^*, g \rangle,
\end{align}
where $f_1, f_2 \in \mathcal{P}$ and $g \in \mathcal{A}$. 
It seems reasonable to conjecture that 
these identities do hold for $n=1$. 
This detail is left to the reader's further consideration. 

For the first property \eqref{comp-prop-1}
the counterexample is provided by taking 
$f_1 = \theta_1$, $f_2 = \theta_1 \theta_2$ and 
$g = \overline{\theta}_2 \theta_1  \overline{\theta}_1$.
Then we have on the one hand that
\begin{equation*}
\langle f_1, f_2 g \rangle \! = \! 
\langle \theta_1 , \theta_1 \theta_2 
\overline{\theta}_2 \theta_1  \overline{\theta}_1 \rangle 
\! = \! w(1,2) \delta_{(1,2)(1,2)} \langle 1, \theta_1  \overline{\theta}_1 \rangle
\! = \! w(1,2) w(1) \ne  0.
\end{equation*}
On the other hand
$$
\langle f_1 g^* , f_2 \rangle = \langle \theta_1 \theta_1 \overline{\theta}_1 \theta_2,
\theta_1 \theta_2 \rangle = 0.
$$
For the second property \eqref{comp-prop-2}, 
we take $f_1 = f_2 = \theta_1$ 
and $g = \theta_2 \overline{\theta}_2$. 
Then we have for the left side that
\begin{equation*}
\langle f_1, f_2 g \rangle = \langle \theta_1, 
\theta_1  \theta_2 \overline{\theta}_2 \rangle
= w(1,2) \delta_{(1,2), (1,2)} \langle 1, 1 \rangle
=  w(1,2) \ne 0.
\end{equation*}
But for the right side we get
\begin{equation*}
\langle f_1  f_2^*, g \rangle = 
\langle \theta_1 \overline{\theta}_1, \theta_2 \overline{\theta}_2 
\rangle 
= w(1,2) \delta_{(1,2), (2,1)} \langle 1, 1 \rangle
= 0. 
\end{equation*}

Thus this example is not compatible with all of the general theory 
presented in \cite{me} when $n\ge 2$. 
But it still is an illuminating example 
as we shall discuss in more detail a bit later on.
However, we do have a particular case of 
(\ref{comp-prop-1}) in this example. 
The only change from (\ref{comp-prop-1}) in the following is that now
$g \in \mathcal{P} \cup \mathcal{P}^*$ is required instead of $g \in \mathcal{A}$. 
\begin{theorem}
\label{thm-2.3}
	Suppose that $f_1, f_2, \in \mathcal{P}$ and 
	$g \in \mathcal{P} \cup \mathcal{P}^*$.
	Then 
	$$
	\langle f_1, f_2 g \rangle = \langle f_1 g^* , f_2 \rangle 
	$$
\end{theorem}
\noindent
\textbf{Proof:}
We first prove the result for $g \in \mathcal{P}$. 
It suffices to consider 
$f_1 = \theta_i$, $f_2 = \theta_j$ and 
$g = \theta_k$ for multi-indices $ i,j,k $. 
Then we get 
\begin{equation*}
\langle f_1, f_2 g \rangle = \langle \theta_i, \theta_j \theta_k \rangle
=  \langle \theta_i, \theta_{(j,k)} \rangle 
= w(i) \delta_{i, (j,k)}. 
\end{equation*}
Next the other side evaluates to 
\begin{equation*}
\langle f_1 g^* , f_2 \rangle = \langle \theta_i (\theta_k)^* , \theta_j \rangle
= \langle \theta_i \overline{\theta}_{k^T} , \theta_j \rangle
= w(i) \delta_{i, (j,k)}, 
\end{equation*}
using $(k^T)^T = k$. 
And so the identity holds in this case.

Next suppose that $g \in \mathcal{P}^*$. 
Then we apply the result of the first case to the
element $g^* \in \mathcal{P}$. 
And that will prove this second case as the reader 
can check by using Theorem~\ref{thm-symmetry}. 
$\quad \blacksquare$

\vskip 0.1cm
Continuing our comments about why this is an illuminating example, let us note
that it satisfies the first seven of the eight properties used for the more general 
theory presented in \cite{me}. 
While it does not satisfy in general the eighth property (that $T_g$ and $T_{g^*}$ are adjoints 
on the domain $\mathcal{P}$ for all $g \in \mathcal{A}$), 
it satisfies the weaker version of this property 
given in Theorem~\ref{weak-adjoint-property} below. 

According to the general theory we have to find a  set $\Phi$ 
which must be a Hamel basis of
$\mathcal{P}$ as well as being an orthonormal set. 
Clearly, the candidate is 
$$
\{  \theta_{i_1} \cdots \theta_{i_r} ~|~ 1 \le i_i \le n, \dots,  1 \le i_r \le n \},
$$
the words in the sub-alphabet 
$\{ \theta_1, \dots, \theta_n \}$. 
And this almost works. 
We need only to normalize these words. 
Taking $f=  \theta_{i_1} \cdots \theta_{i_r}$ 
and $g=\theta_{k_1} \cdots \theta_{k_t}$ in
\eqref{both-lengths-positive} 
(so we have $s=u=0$, $f^\prime =1$ and $g^\prime = 1$) we get 
$$
\langle f , g \rangle = w(i) \delta_{r,t} \delta_{i,k}, 
$$
where $ i = (i_{1} , \dots , i_{r}) $ 
and $ k = (k_{1} , \dots , k_{t}) $ are 
multi-indices of lengths $ r,t $ respectively. 
In particular, $\langle f, g \rangle = 0$ if $f \ne g$. 
On the other hand, $\langle f, f \rangle = w(i) > 0$. 
So we define 
$\varphi_i :=  
w(i)^{-1/2} \, \theta_{i_1} \cdots \theta_{i_r} = 
w(i)^{-1/2} \, \theta_i$, and 
the orthonormal Hamel basis is defined by 
$$
\Phi := \{ \varphi_i   ~|~ 
i = (i_{1}, \dots, i_{r} ) \,\, \mathrm{with} \,\,
r \ge 0, \, 
1 \le i_i \le n, \dots,  1 \le i_r \le n \}.
$$

This argument shows that the complex symmetric 
sesqui-linear form  
restricted to $\mathcal{P}$ is 
positive definite, 
that is, it is an inner product. 
We let $\mathcal{H}$ denote the completion of $\mathcal{P}$ with respect to 
this inner product. 
Then $\Phi$ is an orthonormal basis of $\mathcal{H}$. 
However, 
it is sometimes more convenient to work with the 
orthogonal (but perhaps not orthonormal) set 
$\{ \theta_i = \theta_{i_1} \cdots \theta_{i_r} \}$ of $ \mathcal{H} $. 

We now have enough information about 
the sesqui-linear form in order to define 
the projection $ P : \mathcal{A} \to \mathcal{P} $. 
\begin{definition}
Let $ g \in \mathcal{A} $ be given. 
Then define 
\begin{equation}
\label{define-P}
    P g := \sum_{\varphi_{i} \in \Phi} \langle \varphi_{i} , g \rangle \, \varphi_{i}.
\end{equation}
\end{definition}
This will be well defined when we show in a 
moment that the 
sum on the right side of \eqref{define-P} 
is finite. 
If this were a Hilbert space setting, we could 
write 
$ P = \sum_{i} | \varphi_{i} \rangle \langle \varphi_{i}| $
in Dirac notation, and $ P $ would be 
the orthogonal projection onto the closed 
subspace $ \mathcal{P} $. 
Anyway, this abuse of notation motivates 
the definition of $ P $.  
Given that $ P $ is well-defined, 
it is clear that $ P $ is linear, that it acts 
as the identity on $ \mathcal{P} $ 
(since $ \Phi $ is an 
orthonormal basis of 
$ \mathcal{P} $) and that its
range is $ \mathcal{P} $. 

\begin{theorem}
	The sum on the right side of \eqref{define-P} 
	has only finitely many non-zero terms. 
	Consequently, $ P $ is well-defined. 
\end{theorem}
\noindent 
\textbf{Proof:}
Take $g \in \mathcal{A}$. 
It suffices to show that 
$\langle \theta_i , g \rangle = 0$ except for 
finitely many multi-indices $ i $, since 
$ \varphi_{i} $ is proportional to $ \theta_i $. 
So it suffices to calculate 
$\langle \theta_i , g \rangle$ for all possible 
multi-indices $i$. 
We do this by cases. 

If $g$ begins with a 
$\overline{\theta}$, then 
$\langle \theta_i, g \rangle = 0$ for 
all multi-indices $ i \ne \emptyset$, the 
empty multi-index. 
(Note that $ \theta_{\emptyset} = 1 $.)
It follows that all of the terms, except 
possibly one term, 
on the right side of \eqref{define-P} are $ 0 $
and so $P(g) = \langle 1, g \rangle \, 1$ 
in this case.
For the particular case $ g=1 $ 
(i.e., $ l (g) = 0 $)
we have 
that $P(1) =1$, using
$ \langle 1 , 1 \rangle = 1 $. 

So the only remaining the case is 
when $g$ begins with a $\theta$ and $l(g) > 0$.
Then, using 
$g = \theta_k \overline{\theta}_l g^\prime$ 
for multi-indices $ k,l $ of lengths $ t\ge 1$  
and $u \ge 0 $ 
respectively and $ g^{\prime} $ begins with a 
$ \theta $, we have
\begin{align}
\langle 
\theta_i, g \rangle &= \langle \theta_{i_1} \cdots \theta_{i_r}, 
\theta_{k_1} \cdots \theta_{k_t} \overline{\theta}_{l_1} \cdots  \overline{\theta}_{l_u} g^\prime \rangle 
\label{theta-i-g}
\\
&= w(i,l^T) \, \delta_{r+u,t} \, \delta( (i,l^T), k ) \, \langle 1, g^\prime \rangle 
\nonumber 
\\
&= w(k) \, \delta_{r+u,t} \,
\delta( (i,l^T), k ) \, \langle 1, g^\prime 
\rangle.  
\nonumber 
\end{align} 
Whether this is non-zero now is the question. 
More explicitly, for a given $g$ of this form 
how many $\theta_i$'s 
are there such that this expression could be 
non-zero? 
However, if the factor $\delta( (i,l^T), k )$ is non-zero, then we have necessarily that 
the multi-index $i = (i_{1}, \dots , i_{r} )$ 
of variable length $ r \ge 0 $
forms the initial $r$ entries 
in the given multi-index $k$ of length $ t \ge 1$. 
Thus for a given $ g $ 
there are at most finitely many 
$\theta_i$ for which \eqref{theta-i-g} 
could be non-zero. 
So the sum on the right side of 
\eqref{define-P} has only finitely many 
non-zero terms, and $ P $ is well-defined.  
$ \quad \blacksquare $
\begin{theorem}
$ P $ is symmetric with respect to the 
sesqui-linear form, that is,
$ \langle P f, g \rangle = \langle f, Pg \rangle  $ 
for all $ f,g \in \mathcal{A} $.  
\end{theorem}
\noindent
\textbf{Proof:}
Using 
Theorem~\ref{thm-symmetry} to justify  
the third equality, 
we calculate 
\begin{align*}
\langle 
P f, g \rangle &= 
\big\langle \sum_{i} \langle \varphi_{i} , f \rangle 
\varphi_{i} , g 
\big\rangle 
= 
\sum_{i} \big\langle  
\langle \varphi_{i} , f \rangle 
\varphi_{i} , g 
\big\rangle 
= 
\sum_{i} 
\langle f ,\varphi_{i}  \rangle 
\langle  
\varphi_{i} , g 
\rangle 
\\
&= 
\sum_{i} 
\big\langle f , \langle  
\varphi_{i} , g 
\rangle \varphi_{i}  
\big\rangle = 
\big\langle f , \sum_{i}  \langle  
\varphi_{i} , g 
\rangle \varphi_{i}  
\big\rangle 
=
\langle f , P g \rangle. 
\end{align*}
We also used the finite 
additivity of the sesqui-linear form in each entry,
since the sums have 
only finitely many non-zero terms. 
$ \quad \blacksquare $

\vskip 0.1cm
This result says that $ P $ has {\em an} adjoint 
on $ \mathcal{A} $, namely $ P $ itself. 
Since the sesqui-linear form may be degenerate, 
adjoints need not be unique.

\begin{theorem}
	\label{weak-adjoint-property}
	Suppose that $g \in \mathcal{P} \cup \mathcal{P}^*$. 
	Then for all $f_1, f_2 \in \mathcal{P}$ we have 
	$
	\langle f_1, T_g f_2 \rangle = \langle T_{g^*} f_1 , f_2 \rangle. 
	$
\end{theorem}
\noindent 
\textbf{Proof:} 
Using the previous result and Theorem~\ref{thm-2.3}
we calculate
\begin{align*}
\langle f_{1}, T_{g} f_{2} \rangle 
&= \langle f_{1}, P ( f_{2} g ) \rangle 
= \langle P f_{1}, f_{2} g  \rangle 
= \langle f_{1}, f_{2} g  \rangle 
= \langle f_{1} g^{*}, f_{2} \rangle 
\\
&= \langle f_{1} g^{*}, P f_{2} \rangle 
= \langle P ( f_{1} g^{*} ), f_{2} \rangle 
= \langle T_{g^{*}} f_{1} , f_{2} \rangle. 
\quad \blacksquare
\end{align*}
Since the sesqui-linear form is 
an inner product 
when restricted to $ \mathcal{P} $, 
$  T_{g^*} $ is the
{\em unique} adjoint of 
$ T_{g} $ on $ \mathcal{P} $.
Symmetrically, 
$ T_{g} $ is the 
{\em unique} adjoint of $  T_{g^*} $
on $ \mathcal{P} $. 

Next, for all $\phi \in \mathcal{P}$ we define 
the {\em creation} and {\em annihilation operators} 
associated to the variables $\theta_j$,  $\overline{\theta}_j$ 
for $ 1 \le j \le n $ by 
\begin{equation*}
A_j^\dag (\phi) := T_{\theta_j} (\phi) = P (\phi \theta_j) = \phi \theta_j 
\quad \mathrm{and} \quad 
A_j (\phi) := T_{\overline{\theta}_j} (\phi) = P (\phi \overline{\theta}_j), 
\end{equation*}
respectively. 
These are operators densely defined in $\mathcal{H}$ 
sending $\mathcal{P}$ to itself. 
By Theorem~\ref{weak-adjoint-property}
the operators 
$ A_j^\dag $ and $ A_j $ are adjoints of each other  
on the domain $\mathcal{P}$. 

We now 
evaluate these operators on the basis elements 
$ \varphi_{i} $
of $\mathcal{P}$, where $ i $ is a multi-index.  
First, for the creation operator we have
\begin{align*}
A_j^\dag (\varphi_i) &= P(\varphi_i \theta_j) 
= \varphi_i \theta_j 
=  w(i)^{-1/2} \, \theta_i \, \theta_j 
=  w(i)^{-1/2} \, \theta_{(i,j)}
\\
&= \left( \dfrac{w(i,j)}{w(i)} \right)^{1/2} \!\! \varphi_{(i,j)}. 
\end{align*}

Here $j = 1, \dots, n$ and also $j$ 
denotes the multi-index with exactly one entry,
namely the integer $j$. 
Also we are using the notation $(i,j)$ for 
the multi-index with the integer $j$ concatenated to 
the right of the multi-index $i$. 
It follows that the kernel of $A_j^\dag$ is zero as the reader can check. 
Also, the weight of the `higher' state $\varphi_{(i,j)}$ appears in the numerator 
while the weight of the `lower' state $\varphi_i$ is in the denominator. 
This turns out to be consistent with the way the weights (which are products of factorials) work 
in the case of standard quantum mechanics. 

Next, for the annihilation operator 
$ A_{j} $ for $1 \le j \le n$
we have to evaluate 
$A_j (\varphi_k) = P(\varphi_k\overline{\theta}_j) 
= \sum_{i} \langle \varphi_{i},
\varphi_k\overline{\theta}_j \rangle \varphi_{i} 
$ 
for every multi-index $ k $. 
To do this, consider 
\begin{align*}
&\langle \varphi_i, \varphi_k \overline{\theta}_j \rangle = 
( w(i) w(k) )^{-1/2} \langle \theta_i , \theta_k \overline{\theta}_j \rangle 
\\
&= ( w(i) w(k) )^{-1/2} \, w(k) \, \delta_{r+1, t} \, \delta ( (i,j^T), k ) \, \langle 1, 1 \rangle 
\\
&= \left( \dfrac{w(k)}{w(i)} \right)^{1/2} 
\!\!\! \delta_{r+1, t} \, \delta ( (i,j), k ), 
\end{align*}
where the multi-indices $i = (i_1, \dots, i_r)$ and 
$k = (k_1, \dots, k_t)$ have lengths 
$r$ and $t$ respectively. 
We also used $j^T = j$, since $j$ is a multi-index with exactly one entry in it. 
The only possible non-zero value occurs when the concatenated multi-index 
$(i,j)$ is equal to the multi-index $k$. 
So for $ k \ne (i,j)$ we have that  
$
\langle \varphi_i, \varphi_k \overline{\theta}_j \rangle = 0. 
$
If the last entry in the multi-index $k$ is not $j$  
(i.e., $k_t \ne j$), 
then $ k \ne (i,j)$ for all multi-indices $i$. 
Consequently, in this case we calculate 
$$
A_j (\varphi_k) = P(\varphi_k \overline{\theta}_j ) 
= \sum_i \langle  \varphi_i, \varphi_k \overline{\theta}_j \rangle \varphi_i 
= 0. 
$$

Therefore, 
in this example the annihilation operator $ A_{j} $
has infinite dimensional kernel. 
As a very particular case, we take $k = \emptyset$, the empty multi-index, 
and get that 
$A_{j} ( {\varphi_{\emptyset}}) =
A_j (1) =0$ for all $1 \le j \le n $, 
that is, $ 1 \in \cap_{j=1}^{n} \ker \, A_{j} $. 
So $ 1 $ is a normalized 
vacuum state in $ \mathcal{H} $.  

On the other hand  
if $k= (i,j)$ for some clearly unique 
multi-index $i$ (and in particular $r+1=t$),  
then we find that
$$
\langle \varphi_i, \varphi_{(i,j)} \overline{\theta}_j \rangle = 
\left(
\dfrac{w(i,j)}{w(i)} 
\right)^{1/2} \!\!\! > 0,
$$
and consequently in this case 
$$
A_j (\varphi_{(i,j)} ) = P(\varphi_{(i,j)}  \overline{\theta}_j ) = 
\left(
\dfrac{w(i,j)}{w(i)} 
\right)^{1/2} \!\!\!
\varphi_i. 
$$
Again, the weight of the `higher' state $\varphi_{(i,j)}$ appears in the numerator 
while the weight of the `lower' state $\varphi_i$ is in the denominator. 
And again this is consistent with 
standard quantum mechanics. 

It is now an extended exercise to compute 
the commutation relations of these operators. 
For example, $ [A_{j}^{\dag}, A_{k}^{\dag}] \ne 0 $ 
if $ j \ne k $, since 
$ \theta_{j} \theta_{k} \ne \theta_{k} \theta_{j} $. 
The formulas for these relations 
are simpler if we take 
the weights to be 
$w_i = w_{i_1, \dots, i_r} := 
\mu_{i_1} \cdots \mu_{i_r}$ 
for positive real numbers 
$\mu_1, \dots, \mu_n$.

\section{Concluding Remarks}

I conclude with possibilities 
for future related research concerning
algorithms that manipulate the words 
in the basis of the algebra $ \mathcal{A} $. 

The sesqui-linear form on $\mathcal{A}$ serves 
mainly to 
define the projection operator $P$, which is   crucial in this quantization 
theory. 
Using this,   
creation operators 
tack on a holomorphic variable on the right (up to a weight factor), 
while annihilation operator chop off 
the appropriate holomorphic variable on the right, if present (again up to a weight),  
and otherwise map the word to zero. 
One can define other projection operators that see more deeply into the word, rather 
than looking at only the rightmost part of the word. 
In general each occurrence of $\overline{\theta}_j$ is erased while at the same time 
some corresponding occurrence of $\theta_j$ is also erased. 
The end result is a word with no $\overline{\theta}$'s at all. 
Moreover, if the original word had no $\overline{\theta}$'s to begin with, 
then it will remain unchanged. 
Basically, the projection map is an algorithm that scans a word 
from one end to the other, eliminating all $\overline{\theta}$'s and some $\theta$'s.

There are many such algorithms. 
To give the reader an idea of this, let us consider scanning a word from left to right 
until we hit the first occurrence of $\overline{\theta}_j$ for some $j$. 
We change the word by eliminating this $\overline{\theta}_j$ and the rightmost occurrence
of $\theta_j$ to the left of this $\overline{\theta}_j$, if there is such an occurrence. 
If there is no occurrence of $\theta_j$ to the left, we define $P$ on this word to be zero. 
Otherwise, we continue scanning from our current position in the word looking for the 
next $\overline{\theta}_k$ for some $k$. 
We repeat the same procedure. 
Since the word is finite in length, this algorithm will terminate. 
At such time there will be no occurrences of  $\overline{\theta}$'s left. 
The resulting word (or zero) will be $P$ evaluated on the original word. 

The reader is invited to produce other algorithms for finding one (or various) occurrences of 
$\theta_j$ to pair with an identified occurrence of $\overline{\theta}_j$. 
There are other deterministic algorithms for sure, but there are even stochastic algorithms as well. 
These stochastic algorithms could pair a random number of occurrences of $\theta_j$, including 
zero occurrences with non-zero probability, with a given occurrence of $\overline{\theta}_j$. 
Also, the locations of these occurrences could be random. 
Then all Toeplitz operators, including those of creation and annihilation, would become 
random operators. 

\vskip 0.2cm
\begin{center}
	{\small ACKNOWLEDGMENT}
\end{center}
\vskip 0.1cm
I thank the organizers for the opportunity 
to participate 
in the event  
Operator Algebras, 
Toeplitz Operators and Related Topics, 
(OATORT) held in Boca del Rio, 
Veracruz, Mexico in
November, 2018.

\end{document}